\documentclass[prl,superscriptaddress,showpacs,twocolumn]{revtex4}

\usepackage{graphics}
\usepackage{amsmath}
\usepackage{feynmp}
\usepackage{slashed}

\renewcommand{\Re}{\mathop{\rm Re}\nolimits}% symbols for the real
\renewcommand{\Im}{\mathop{\rm Im}\nolimits}% and imaginary parts
\newcommand{\Tr}{\mathop{\rm Tr}\nolimits}% trace operator

\begin{document}
\begin{fmffile}{fig}

\title{A model of flavors}
\author{Tom\'a\v{s} Brauner}
\email{brauner@ujf.cas.cz}
\affiliation{Department of Theoretical Physics, Nuclear Physics Institute, 25068 \v Re\v z~(Prague), Czech Republic}
\affiliation{Faculty of Mathematics and Physics, Charles University, Prague, Czech Republic}
\author{Ji\v{r}\'{\i} Ho\v{s}ek}
\affiliation{Department of Theoretical Physics, Nuclear Physics Institute, 25068 \v Re\v z~(Prague), Czech Republic}

\begin{abstract}
We argue in favor of dynamical mass generation in an $SU(2)_L\times U(1)_Y$
electroweak model with two complex scalar doublets with ordinary masses. The
masses of leptons and quarks are generated by ultraviolet-finite
non-perturbative solutions of the Schwinger--Dyson equations for full fermion
propagators with self-consistently modified scalar boson exchanges. The $W$ and
$Z$ boson masses are expressed in terms of spontaneously generated fermion
proper self-energies in the form of sum rules. The model predicts two charged
and four real neutral heavy scalars.
\end{abstract}

\maketitle

The standard model of electroweak interactions is the best what in theoretical
particle physics we have: In an operationally well defined framework it
parameterizes and successfully correlates virtually all electroweak phenomena.
Yet as an effective field theory it is ill with unnaturalness
\cite{Polchinski:1992ed}: Due to the necessity of its quadratic renormalization
there is no natural way how the Higgs mass could be made reasonably small.
Natural ways end up with the Higgs mass uninterestingly high, of the order of
the cutoff $\Lambda\approx10^{16}\,\text{GeV}$.

Possibilities of curing unnaturalness we are aware of are not numerous. First,
supersymmetry models avoid the quadratic divergences associated with scalar
fields by extending the field spectrum to become fermion-boson symmetric.
Second, the `Little Higgs' models (see \cite{Schmaltz:2002wx} for introduction)
avoid quadratic divergences associated with the Higgs boson mass at low
energies by introducing new symmetries which generate sophisticated mixings
between bosons. Third, the models without elementary scalar fields
\cite{Weinberg:1979bn,Susskind:1979ms} avoid the quadratic divergences by
assumption. Unlike the first two possibilities they are necessarily strongly
coupled and, in fact, they are not operationally well defined.

With the ultimate aim of generating the particle masses softly and with not
vastly different couplings we suggest in this note to add to the list of models
of spontaneous mass generation a~dynamical one within an effective field theory
description of the electroweak phenomena: Massive scalar fields distinguish
fermions at tree level by their different $SU(2)_L\times U(1)_Y$ invariant
Yukawa couplings, while spontaneous breakdown of this symmetry i.e., emergence
of both the fermion and the intermediate-boson masses is a~genuinely quantum
self-consistent loop effect. This alternative is arguably also natural.

In a~self-explanatory notation our $SU(2)_L \times U(1)_Y$ gauge-invariant
electroweak model is defined by its Lagrangian ${\cal L}$ which consists of (i)
Four standard massless gauge fields; (ii) $n_f$ standard massless fermion
families extended by $n_f$ neutrino right-handed singlets with zero weak
hypercharge together with their general Majorana mass matrix $M_M$; (iii) two
distinct complex scalar doublets $S = (S^{(+)},S^{(0)})$ and $N =
(N^{(0)},N^{(-)})$ with weak hypercharges $Y(S)=+1$ and $Y(N)=-1$,
respectively, with different $\it{ordinary}$ masses $M_S$ and $M_N$,
respectively, and with different respective self-couplings $\lambda_S,
\lambda_N$. With spontaneous breakdown of the underlying symmetry in the scalar
sector at tree level such a~Lagrangian would define a~popular extension of the
Standard model \cite{Weinberg:1976hu}.

The scalar field doublets are of utmost importance for the `low-energy' physics
of electroweak symmetry breaking. Their Yukawa couplings, which we assume to
have the form
\begin{multline}
{\cal L}_Y=\bar l_Ly_ee_RS+\bar l_Ly_{\nu}\nu_RN+\\
+\bar q_Ly_dd_RS+\bar q_Ly_uu_RN+\text{H.c.},
\label{Yukawa_int}
\end{multline}
distinguish between otherwise identical fermion families, and break down
explicitly all unwanted and dangerous inter-family symmetries. Both $M_M$ and
$S, N$ together with their Yukawa couplings are thought of as remnants of an
unknown high-energy dynamics. Because scalars ought to remain in the physical
spectrum, and the neutral ones mediate in general the flavor-changing electric
charge conserving transitions, for safety reasons we restrict their masses as
$M_S,M_N\gtrsim10^6\,\text{GeV}$.

The troublesome quadratic renormalizations of $M_S, M_N$ can easily be avoided
at will by imposing mass-independent relations between the gauge couplings $g,
g'$, the Yukawa couplings $y$, and the scalar self couplings $\lambda$
\cite{Veltman:1981mj}. The Yukawa couplings $y_e, y_{\nu},y_d, y_u$ are in
general arbitrary complex constant $n_f\times n_f$ matrices unconstrained by
the symmetry. The Lagrangian ${\cal L}$ is at least one-loop renormalizable off
mass shell by power counting, and all its counter terms are only
logarithmically divergent. At very high momenta at which all dynamically
generated masses can be neglected it is possible to diagonalize two general
complex matrices, say $y_{\nu}$ and $y_u$, by biunitary transformations into
the real non-negative form without changing physics.

It should, however, be the dynamical issue of small momenta, what are the
observable masses of leptons, quarks, and intermediate vector bosons.

I. Our intention is to break down the sacred $SU(2)_L \times U(1)_Y$ symmetry
dynamically. This amounts to first \emph{assuming}, and then self-consistently
\emph{finding} non-zero fermion proper self-energies $\Sigma(p)$ due to the
interactions. Their chirality-changing part must come out necessarily
ultraviolet-finite because the fermion mass counter terms are strictly
forbidden by the underlying chiral $SU(2)_L \times U(1)_Y$ symmetry. Like in
the Standard model we assume also here that the fermion--gauge boson
interactions are not the cause of the fermion masses, and they will therefore
not be considered in this respect. The assumption is of course natural because
these interactions do not feel flavors. Like in the Standard model the fermion
masses will be generated by the Yukawa couplings of fermions with scalar
doublets which do feel flavors. Unlike the Standard model the fermion masses
will bona fide be the genuinely quantum loop effect.

We will now show how the fermion masses can be generated by means of 
a~self-consistent solution to the Schwinger--Dyson equations. For simplicity we
will, in the case of neutrinos, consider just the Dirac masses. Later we shall
discuss how the Majorana mass of the right-handed neutrino affects the present
mechanism.

\begin{figure}
$$
{\def\figlab{e}% local definition of label for fermion lines
\parbox{30\unitlength}{%
\begin{fmfgraph*}(30,15)
\fmfkeep{loop}
\fmfset{arrow_len}{0.1w}
\fmfleft{l}
\fmfright{r}
\fmf{scalar,label=$S^{(0)}$,l.si=left,tension=4}{vl,l}
\fmf{scalar,label=$S^{(0)}$,l.si=right,tension=4}{vr,r}
\fmf{phantom,right}{vr,vl}
\fmf{phantom,right}{vl,vr}
\fmffreeze
\fmfipath{p[]}
\fmfiset{p1}{vpath(__vr,__vl)}
\fmfiset{p2}{vpath(__vl,__vr)}
\fmfiv{d.sh=circle,d.si=0.1w,d.fi=full}{point length(p1)/2 of p1}
\fmfiv{d.sh=circle,d.si=0.1w,d.fi=full}{point length(p2)/2 of p2}
\fmfi{fermion,label=$\figlab_R$,l.si=right}{subpath(0,length(p1)/2) of p1}
\fmfi{fermion,label=$\figlab_L$,l.si=right}{subpath(length(p1)/2,length(p1)) of p1}
\fmfi{fermion,label=$\figlab_R$,l.si=right}{subpath(0,length(p2)/2) of p2}
\fmfi{fermion,label=$\figlab_L$,l.si=right}{subpath(length(p2)/2,length(p2)) of p2}
\end{fmfgraph*}}% end of parbox
\,+\,
\def\figlab{d}
\parbox{30\unitlength}{\fmfreuse{loop}}}
$$
\caption{One-loop graphs that induce mixing of the real and imaginary
components of the neutral scalar fields, leading to their mass splitting. The
solid circles denote the chirality-changing part of the fermion self-energy
$\Sigma(p)$. The same graphs also apply to $N^{(0)}$ upon replacing $e,d$ with
$\nu,u$.}
\label{Fig:scalar_mixing}
\end{figure}
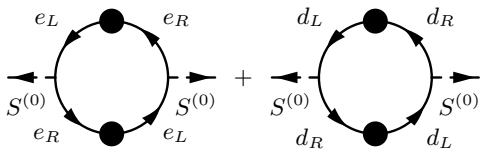
II. Simple but crucial is the observation that the assumed fermion mass
insertions give rise to generically new contributions to the propagators of the
scalar fields. Let us first consider the electrically neutral complex scalars
$S^{(0)}$ and $N^{(0)}$. As shown in Fig. \ref{Fig:scalar_mixing}, the fermion
masses induce non-zero two-point functions $\langle S^{(0)}S^{(0)}\rangle$ and
$\langle N^{(0)}N^{(0)}\rangle$. We write down explicitly the corresponding
proper self-energy (and the following formulas) for the `southern' scalar,
\begin{multline*}
\mu^2_S(p^2)\equiv-i\int\frac{d^4k}{(2\pi)^4}\Tr\Bigl[y_e^{\dagger}
G_e^{LR}(k)y_e^{\dagger}G_e^{LR}(k+p)\Bigr]-\\
-i\int\frac{d^4k}{(2\pi)^4}\Tr\Bigl[y_d^{\dagger}
G_d^{LR}(k)y_d^{\dagger}G_d^{LR}(k+p)\Bigr],
\end{multline*}
where the superscript $LR$ denotes that part of the full propagator $G$, which
in the chiral basis connects the left- and right-handed components of the Dirac
field.

Since the scalar mass $M_S$ is expected to change only slightly by the
interaction, we may for simplicity evaluate $\mu^2_S(p^2)$ at $p^2=M_S^2$. We
also neglect the one-loop renormalization of the `ordinary' mass $M_S$, which
can be justified by a~suitable choice of the renormalization prescription. The
problem thus reduces to finding the spectrum of the bilinear Lagrangian
\begin{multline*}
{\cal L}^{(0)}_S=\partial_{\mu}S^{(0)}\partial^{\mu}S^{(0)\dagger}-M_S^2S^{(0)}S^{(0)\dagger}-\\
-\frac12\mu_S^{2*}S^{(0)}S^{(0)}-\frac12\mu_S^2S^{(0)\dagger}S^{(0)\dagger}.
\end{multline*}

Physically observable are then two real spin-0 particles. Their masses are
$$
M_{1,2S}^2=M_S^2\pm\left|\mu_S^2\right|.
$$
The corresponding real fields $S_1$ and $S_2$ are defined through
$$
S^{(0)}=\frac1{\sqrt2}e^{i\alpha_S}(S_1+iS_2),
$$
where the `mixing angle' $\alpha_S$ is merely given by the phase of the
anomalous mass term, $\tan2\alpha_{S}=\Im\mu_S^2/\Re\mu_S^2$. All formulas
written down explicitly for $S$ hold of course also for $N$.

This is to the mixing of the neutral scalars at the one-loop level. One should,
however, note that at higher orders (and, possibly, upon switching on the gauge
interaction) we also find transitions between the `southern' and `northern'
scalars. Consequently, there are altogether four real scalars that mix with one
another, and they in fact should because there is no symmetry that would
prohibit the mixing.

\begin{figure}
$$
{\def\figlabu{\nu}% local definitions of labels for fermion lines
\def\figlabd{e}
\parbox{30\unitlength}{%
\begin{fmfgraph*}(30,15)
\fmfkeep{loop_charged}
\fmfset{arrow_len}{0.1w}
\fmfleft{l}
\fmfright{r}
\fmf{scalar,label=$S^{(+)}$,l.si=left,tension=4}{vl,l}
\fmf{scalar,label=$N^{(-)}$,l.si=right,tension=4}{vr,r}
\fmf{phantom,right}{vr,vl}
\fmf{phantom,right}{vl,vr}
\fmffreeze
\fmfipath{p[]}
\fmfiset{p1}{vpath(__vr,__vl)}
\fmfiset{p2}{vpath(__vl,__vr)}
\fmfiv{d.sh=circle,d.si=0.1w,d.fi=full}{point length(p1)/2 of p1}
\fmfiv{d.sh=circle,d.si=0.1w,d.fi=full}{point length(p2)/2 of p2}
\fmfi{fermion,label=$\figlabu_R$,l.si=right}{subpath(0,length(p1)/2) of p1}
\fmfi{fermion,label=$\figlabu_L$,l.si=right}{subpath(length(p1)/2,length(p1)) of p1}
\fmfi{fermion,label=$\figlabd_R$,l.si=right}{subpath(0,length(p2)/2) of p2}
\fmfi{fermion,label=$\figlabd_L$,l.si=right}{subpath(length(p2)/2,length(p2)) of p2}
\end{fmfgraph*}}% end of parbox
\,+\,
\def\figlabu{u}
\def\figlabd{d}
\parbox{30\unitlength}{\fmfreuse{loop_charged}}}
$$
\caption{One-loop amplitude for the charged scalar mixing induced by the Yukawa
interaction.}
\label{Fig:charged_scalar_mixing}
\end{figure}
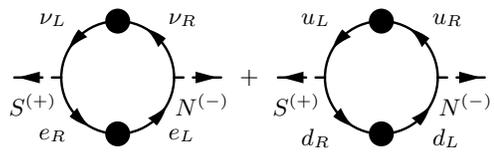
The case of the charged scalars is considerably simpler. Now only particles
with the same charge can mix and they really do as shown in Fig.
\ref{Fig:charged_scalar_mixing}. Analogously to the previous case, we introduce
a~mixing term
$$
-\mu_{SN}^{2*}S^{(+)}N^{(-)}-\mu_{SN}^2S^{(+)\dagger}N^{(-)\dagger}
$$
into the Lagrangian.

The spectrum then contains two charged complex scalars, referred to as
$\Phi_1^{(+)}$ and $\Phi_2^{(+)}$, that are mixtures of $S^{(+)}$ and
$N^{(-)\dagger}$. Their masses are given by
$$
M_{1,2\Phi}^2=\frac12\left[M_S^2+M_N^2\pm
\sqrt{(M_S^2-M_N^2)^2+4\left|\mu^2\right|^2}\right]
$$
and the corresponding field transformation is
\begin{align*}
S^{(+)}&=e^{+i\alpha_{SN}}\left(\Phi_1^{(+)}\cos\theta-\Phi_2^{(+)}\sin\theta\right),\\
N^{(-)\dagger}&=e^{-i\alpha_{SN}}\left(\Phi_1^{(+)}\sin\theta+\Phi_2^{(+)}\cos\theta\right),
\end{align*}
where $\alpha_{SN}$ is the phase of $\mu_{SN}$ and
$$
\tan2\theta=\frac{2\left|\mu_{SN}^2\right|}{M_S^2-M_N^2}.
$$

The splittings $\mu_S^2, \mu_N^2$ and $\mu_{SN}^2$ of the scalar-particle
masses due to the (yet assumed) dynamical fermion mass generation are both
natural and important: First, they come out UV finite due to the large-momentum
behavior of $\Sigma(p^2)$ (see below). Second, they manifest spontaneous
breakdown of the $SU(2)_L \times U(1)_Y$ symmetry down to $U(1)_{em}$ in the
scalar sector. Third, they will be responsible for the necessary ultraviolet
finiteness of both the fermion and the intermediate vector boson masses.

\begin{figure}
$$
{\def\figlab{$S_1$}% local definition of label for scalar lines
\parbox{30\unitlength}{%
\begin{fmfgraph*}(30,25)
\fmfkeep{sunrise}
\fmfset{arrow_len}{0.1w}
\fmfleft{l}
\fmfright{r}
\fmf{fermion,label=$e_R$,l.si=left,tension=4}{r,vr}
\fmf{fermion,label=$e_L$,l.si=left,tension=4}{vl,l}
\fmf{dashes,right,label=\figlab,l.si=right}{vr,vl}
\fmf{fermion,label=$e_L$,l.si=left}{vr,v}
\fmf{fermion,label=$e_R$,l.si=left}{v,vl}
\fmfv{d.sh=circle,d.si=0.1w,d.fi=full}{v}
\end{fmfgraph*}}% end of parbox
\,-\,
\def\figlab{$S_2$}
\parbox{30\unitlength}{\fmfreuse{sunrise}}
}\,+
$$
$$
+\,{\def\figlab{$\Phi_1^{(+)}$}% local definition of label for scalar lines
\parbox{30\unitlength}{%
\begin{fmfgraph*}(30,15)
\fmfkeep{sunrise_charged}
\fmfset{arrow_len}{0.1w}
\fmfleft{l}
\fmfright{r}
\fmf{fermion,label=$e_R$,l.si=left,tension=4}{r,vr}
\fmf{fermion,label=$e_L$,l.si=left,tension=4}{vl,l}
\fmf{scalar,left,label=\figlab,l.si=left}{vl,vr}
\fmf{fermion,label=$\nu_L$,l.si=left}{vr,v}
\fmf{fermion,label=$\nu_R$,l.si=left}{v,vl}
\fmfv{d.sh=circle,d.si=0.1w,d.fi=full}{v}
\end{fmfgraph*}}% end of parbox
\,-\,
\def\figlab{$\Phi_2^{(+)}$}
\parbox{30\unitlength}{\fmfreuse{sunrise_charged}}
}
$$
\caption{One-loop contributions to the chirality-changing part of the electron
proper self-energy. The minus signs suggest that we encounter the difference of
the two scalar propagators. The same graphs also apply to $d$-quarks, and (upon
replacing $S_{1,2}$ with $N_{1,2}$ and reversing the arrows on the charged
scalar propagators) neutrinos and $u$-quarks.}
\label{Fig:Schwinger_Dyson}
\end{figure}
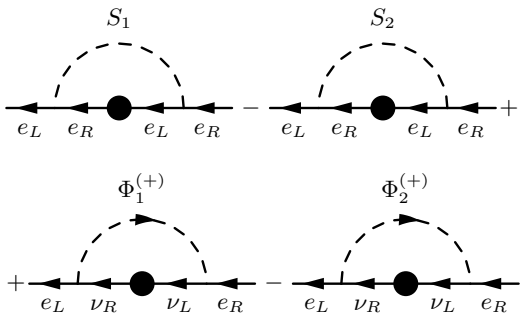
III. In a~self-consistent one-loop approximation the Yukawa interactions
(\ref{Yukawa_int}) give rise to the Schwinger--Dyson equation for the matrix
fermion proper self-energy. Since we are interested in symmetry breaking
solutions of the Schwinger--Dyson equation, we use the following
simplification: We abandon the symmetry-preserving part of the radiative
corrections to the fermion propagators that is, we neglect the wave function
renormalization \cite{Cornwall:1974vz}. The fermion self-energy thus reduces to
its chirality changing parts $\Sigma^{LR}$ and $\Sigma^{RL}$. To simplify the
notation, we set $\Sigma\equiv\Sigma^{LR}$.

The Schwinger--Dyson equation then acquires the form, depicted in Fig.
\ref{Fig:Schwinger_Dyson} explicitly for the case of charged leptons:
\begin{widetext}
\begin{multline}
\label{SD_equation}
\Sigma_e(p^2)=\frac i2\int\frac{d^4k}{(2\pi)^4}\left\{e^{2i\alpha_S}y_e
\Sigma_e^{\dagger}(k^2)\bigl[k^2-\Sigma_e(k^2)\Sigma_e^{\dagger}(k^2)\bigr]^{-1}y_e
\left[\frac1{(p-k)^2-M_{1S}^2}-\frac1{(p-k)^2-M_{2S}^2}\right]\right.+\\
+\left.e^{2i\alpha_{SN}}\sin2\theta y_{\nu}\Sigma_{\nu}^{\dagger}(k^2)
\bigl[k^2-\Sigma_{\nu}(k^2)\Sigma_{\nu}^{\dagger}(k^2)\bigr]^{-1}y_e
\left[\frac1{(p-k)^2-M_{1\Phi}^2}-\frac1{(p-k)^2-M_{2\Phi}^2}\right]\right\}.
\end{multline}
\end{widetext}

Worth of mentioning is that the assumed fermion mass generation results in the
scalar boson mixings which \emph{guarantee} that the kernel of the
Schwinger--Dyson equation is Fredholm \cite{Pagels:1980ai}. In order to proceed
we are at the moment forced to resort to simplifications: We neglect the
fermion mixing and set $\sin2\theta=0$. This amounts formally to neglecting the
charged boson mixing. More important, it physically amounts to neglecting an
interesting relation between the masses of the up ($U$) and the down ($D$)
fermions in a~weak fermion doublet. With these simplifications, keeping the form
of the nonlinearity unchanged, we perform in Eq. (\ref{SD_equation}) the Wick
rotation, angular integrations, and Taylor expand in $M_{1S}^2-M_{2S}^2$. For 
a~generic (say $e$) fermion self-energy in dimensionless variables $\tau=p^2/M^2$
we obtain
\begin{equation}
\Sigma(\tau)=\beta\int_0^\infty d\sigma\frac{\sigma \Sigma(\sigma)}
{\sigma +\frac{1}{M^2} \Sigma^2(\sigma)}\kappa(\sigma,\tau),
\label{integral_eq}
\end{equation}
where
$$
\kappa(\sigma,\tau)\equiv\frac{[(\sigma+\tau+1)^2-4\sigma\tau]^{-1/2}}
{{\sigma+\tau+1+[(\sigma+\tau+1)^2-4\sigma\tau]^{1/2}}}
$$
and $\beta=(y^2/8\pi^2)(M_1^2-M_2^2)/M^2$. Here $M$ is the common mean mass
about which $M_1$ and $M_2$ are expanded. Numerical analysis of the Eq.
(\ref{integral_eq}) reveals the existence of a~solution $\Sigma(\tau)$, albeit
yet for large values of $\beta$ \cite{Pagels:1980ai}. It has a~form similar to
the step function rapidly approaching zero after the step at $\tau=1$. It
correctly exhibits the low-momentum origin of the fermion masses. The model can
pretend to phenomenological relevance only after demonstrating strong
amplification of fermion masses as a~response to small changes of preferably
small Yukawa couplings. This work is in progress.

For the electrically charged fermions $\Sigma(p^2)$ defines the fermion mass
$m$ as the solution of the equation $m=\Sigma(p^2=m^2)$. The case of neutrinos
is more subtle and requires further work: (i) Without introducing $\nu_R$
neutrinos would be massless in the present model. (ii) With $\nu_R$ the
mechanism described above generates the UV-finite Dirac neutrino self-energy
$\Sigma_{\nu}$. Moreover, there is also a~`hard' mass term present in ${\cal
L}$,
\begin{equation}
\overline{(\nu_R)^{{\cal C}}}M_M\nu_R+\text{H.c.}
\label{M_M}
\end{equation}
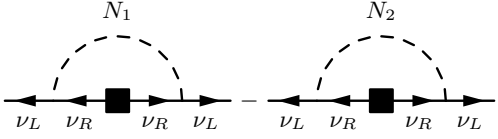
\begin{figure}
$$
{\def\figlab{$N_1$}% local definition of label for scalar lines
\parbox{30\unitlength}{%
\begin{fmfgraph*}(30,10)
\fmfkeep{Majorana}
\fmfset{arrow_len}{0.1w}
\fmfleft{l}
\fmfright{r}
\fmf{fermion,label=$\nu_L$,l.si=right,tension=4}{vr,r}
\fmf{fermion,label=$\nu_L$,l.si=left,tension=4}{vl,l}
\fmf{dashes,right,label=\figlab,l.si=right}{vr,vl}
\fmf{fermion,label=$\nu_R$,l.si=right}{v,vr}
\fmf{fermion,label=$\nu_R$,l.si=left}{v,vl}
\fmfv{d.sh=square,d.si=0.1w,d.fi=full}{v}
\end{fmfgraph*}}% end of parbox
\,-\,
\def\figlab{$N_2$}
\parbox{30\unitlength}{\fmfreuse{Majorana}}
}
$$
\caption{One-loop contributions to the Majorana mass of the left-handed
neutrinos. The solid square denotes the right-handed Majorana mass.}
\label{Fig:Majorana_mass}
\end{figure}%
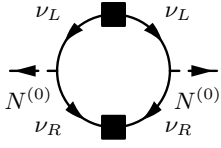
\begin{figure}
$$
\parbox{30\unitlength}{%
\begin{fmfgraph*}(30,15)
\fmfset{arrow_len}{0.1w}
\fmfleft{l}
\fmfright{r}
\fmf{scalar,label=$N^{(0)}$,l.si=left,tension=4}{vl,l}
\fmf{scalar,label=$N^{(0)}$,l.si=right,tension=4}{vr,r}
\fmf{phantom,right}{vr,vl}
\fmf{phantom,right}{vl,vr}
\fmffreeze
\fmfipath{p[]}
\fmfiset{p1}{vpath(__vr,__vl)}
\fmfiset{p2}{vpath(__vl,__vr)}
\fmfiv{d.sh=square,d.si=0.1w,d.fi=full}{point length(p1)/2 of p1}
\fmfiv{d.sh=square,d.si=0.1w,d.fi=full}{point length(p2)/2 of p2}
\fmfi{fermion,label=$\nu_L$,l.si=left}{subpath(length(p1)/2,0) of p1}
\fmfi{fermion,label=$\nu_L$,l.si=right}{subpath(length(p1)/2,length(p1)) of p1}
\fmfi{fermion,label=$\nu_R$,l.si=right}{subpath(0,length(p2)/2) of p2}
\fmfi{fermion,label=$\nu_R$,l.si=left}{subpath(length(p2),length(p2)/2) of p2}
\end{fmfgraph*}}% end of parbox
$$
\caption{One-loop contribution to the $N^{(0)}-N^{(0)\dagger}$ mixing arising
from the neutrino Majorana mass term.}
\label{Fig:scalar_mixing_Majorana}
\end{figure}%
Finally, due to (\ref{M_M}) the model is in general capable of generating 
a~UV-finite left-handed Majorana mass matrix, the mechanism being essentially the
same as that for the Dirac masses, see Fig. \ref{Fig:Majorana_mass}. Strictly
speaking, it is more appropriate to treat the Majorana masses
self-consistently, on the same footing as the Dirac masses. The mixing
amplitude of the neutral `northern' scalars then receives an additional
contribution shown in Fig. \ref{Fig:scalar_mixing_Majorana}.

Consequently, with $\nu_R$ the model should describe $2n_f$ massive Majorana
neutrinos with a~generic see-saw spectrum.

IV. Dynamically generated fermion proper self-energies $\Sigma(p^2)$ break
spontaneously the $SU(2)_L\times U(1)_Y$ symmetry down to $U(1)_{em}$.
Consequently, the $W$ and $Z$ bosons dynamically acquire masses. To determine
their values we have to calculate residues at single massless poles of their
polarization tensors \cite{Schwinger:1962tp}.

(i) The massless poles are those of the `would-be' Nambu--Goldstone (NG)
bosons. They are visualized in the proper vertex functions
$\Gamma^{\alpha}_{W}$ and $\Gamma^{\alpha}_{Z}$ as necessary consequences of
the Ward--Takahashi identities \cite{Margolis:1984cs}:
\begin{multline*}
\Gamma^{\alpha}_{W}(p+q,p)\xrightarrow[q^2\to0]{}
\frac{g}{2\sqrt2}\{\gamma^{\alpha}(1-\gamma_5)-\\
-\frac{q^{\alpha}}{q^2}[(1-\gamma_5)\Sigma_U(p+q)-(1+\gamma_5)\Sigma_D(p)]\},
\end{multline*}
\begin{multline*}
\Gamma^{\alpha}_{Z}(p+q,p)\xrightarrow[q^2\to0]{}
\frac{g}{2\cos\theta_W}\{t_3\gamma^{\alpha}(1-\gamma_5)-\\
-2Q\gamma^{\alpha}\sin^2\theta_W+\frac{q^{\alpha}}{q^2}t_3
[\Sigma(p+q)+\Sigma(p)]\gamma_5\}.
\end{multline*}

(ii) From the pole terms in $\Gamma_W^{\alpha}$ and $\Gamma_Z^{\alpha}$ we
extract the effective vertices between the gauge and three
\emph{multi-component} `would-be' Nambu--Goldstone bosons. They are given in
terms of the UV-finite tadpole loop integrals
\begin{widetext}
\begin{align*}
J_W^{\alpha}(q)&=\Tr\int\frac{d^4k}{(2\pi)^4}P_{-}G_{U}(k+q)
\frac{g}{2\sqrt2}\gamma^{\alpha}(1-\gamma_5)G_{D}(k)
=\frac{g}{\sqrt2}\frac1{N_W}[I^{\alpha}_{U;D}(q)+I^{\alpha}_{D;U}(q)],\\
J_Z^{\alpha}(q)&=\Tr\int\frac{d^4k}{(2\pi)^4}P_0G(k+q)\frac{g}{2\cos\theta_W}
\left[t_3\gamma^{\alpha}(1-\gamma_5)-2Q\gamma^{\alpha}\sin^2\theta_W\right]G(k)
=\frac{g}{2\cos\theta_W}\frac1{N_Z}[I^{\alpha}_{U;U}(q)+I^{\alpha}_{D;D}(q)],
\intertext{where}
I^{\alpha}_{U;D}(q)&\equiv4n_c\int\frac{d^4k}{(2\pi)^4}\frac{\Sigma_U^2(k+q)k^{\alpha}}
{[(k+q)^2-\Sigma^2_U(k+q)][k^2-\Sigma^2_D(k)]}\equiv-iq^{\alpha}I_{U;D}(q^2).
\end{align*}
\end{widetext}
Also $G(k)=[\slashed{k}+\Sigma(k)]/[k^2-\Sigma^2(k)]$,
$P_{\pm}=N_W^{-1}[(1\mp\gamma_5)\Sigma_U(p+q)-(1\pm\gamma_5)\Sigma_D(p)]$,
$P_0=N_Z^{-1}\gamma_5t_3[\Sigma(p+q)+\Sigma(p)]$, and $n_c=1$ for leptons and
$n_c=3$ for quarks. The normalization factors $N_W$, $N_Z$ are defined by the
mass sum rules below. In the loop integrals summation over all families of $U$
and $D$ fermions (both leptons and quarks) is implied.

(iii) The effective gauge-boson--`would-be' NG vertices immediately give rise
to the longitudinal parts of $W$ and $Z$ polarization tensors with massless
`would-be' NG poles. Their residues are
\begin{multline*}
m_W^2=\frac{1}{4}g^2\sum(I_{U;D}(0)+I_{D;U}(0))\equiv\frac{1}{4}g^2N_W^2,\\
m_Z^2=\frac{1}{4}(g^2+g'^2)\sum(I_{U;U}(0)+I_{D;D}(0))\equiv\frac{1}{4}(g^2+g'^2)N_Z^2.
\end{multline*}

If the proper self-energies $\Sigma_U$ and $\Sigma_D$ were degenerate, not
surprisingly the Standard model relation $m_W^2/m_Z^2\cos^2\theta_W$ would be
fulfilled. Quantitative analysis of departure from this relation demands
quantitative knowledge of the functional form of the proper self-energies. At
present we can only refer to an illustrative model analysis of Ref.
\cite{hosek:1987gf}.

V. Generating the lepton, quark, and vector boson masses spontaneously is 
a~theoretical necessity. Principles are, however, more general then their known
realizations. Being genuinely quantum and non-perturbative the mechanism of
spontaneous mass generation suggested here is rather stiff:

(i) It relates all quark (lepton) dynamically generated proper self-energies
$\Sigma(p^2)$ with each other. After some labor the relations between fermion
self-energies should turn into relations between the quark (lepton) masses,
their corresponding mixing angles and the CP-violating phase(s). (ii) It
relates all dynamically generated proper self-energies $\Sigma(p^2)$ to $m_W$
and $m_Z$. (iii) There is no generic weak-interaction mass scale
$v\simeq246\,\text{GeV}$ in the present model. The mass scale of the world is
fixed by the masses $M_N$, $M_S$ and $M_M$.

We believe that with new experimental data soon to come the model might provide
a~slit into yet unknown short-distance particle dynamics. Be it as it may it
predicts four electrically neutral scalar bosons and two charged ones. They
should be heavy, but not too much.

\begin{acknowledgments}
We are grateful to Petr Bene\v{s} for numerically demonstrating the existence
of a~solution of Eq. (\ref{integral_eq}). The present work was supported in
part by grant GACR No. 202/02/0847 and by the ASCR project No. K1048102.
\end{acknowledgments}

\end{fmffile}
\end{document}